\documentclass[11pt]{article}

\usepackage{setspace}

\usepackage{graphicx}

\usepackage{times}
\usepackage{amsmath}
\usepackage{amsfonts}

\usepackage{hyperref}

\usepackage[mathlines]{lineno}

\newenvironment{sciabstract}{%
\begin{quote} \bf}
{\end{quote}}

\newcounter{lastnote}

\title{Geometrical dynamics of edge-driven surface growth}

\author{
C. Nadir Kaplan$^{1, 2\, \ast}$ and L. Mahadevan$^{3-5\, \dagger}$
\\
\normalsize{$^{1}$Department of Physics,}\\
\normalsize{Virginia Polytechnic Institute and State University, Blacksburg, VA 24061, USA.}\\
\normalsize{$^{2}$Center for Soft Matter and Biological Physics,}\\ 
\normalsize{Virginia Polytechnic Institute and State University, Blacksburg, VA 24061, USA.}\\
\normalsize{$^{3}$School of Engineering and Applied Sciences,}\\ 
\normalsize{Harvard University, Cambridge, MA 02138, USA.}\\
\normalsize{$^{4}$Department of Physics, Harvard University, Cambridge, MA 02138, USA.}\\
\normalsize{$^{5}$Department of Organismic and Evolutionary Biology,}\\ 
\normalsize{Harvard University, Cambridge, MA 02138, USA.}\\

\normalsize{$^\ast$ E-mail: nadirkaplan@vt.edu}\\
\normalsize{$^\dagger$E-mail:  lmahadev@g.harvard.edu.}

}

\date{}

\begin{document} 




\maketitle

\begin{sciabstract}
Accretion of mineralized thin wall-like structures via localized growth along their edges is observed in a range of physical and biological systems ranging from molluscan and brachiopod shells to carbonate-silica composite precipitates. To understand the shape of these mineralized structures, we develop a mathematical framework that treats the thin-walled shells as a smooth surface left in the wake of the growth front that can be described as an evolving space curve. Our theory then takes an explicit geometric form for the prescription of the velocity of the growth front curve, along with some compatibility relations and a closure equation related to the nature of surface curling. The result is a set of equations for the geometrical dynamics of a curve that leaves behind a compatible surface. Solutions of these equations capture a range of geometric precipitate patterns seen in abiotic and biotic forms across scales. In addition to providing a framework for the growth and form of these thin-walled morphologies, our theory suggests a new class of dynamical systems involving moving space curves that are compatible with non-Euclidean embeddings of surfaces.
\end{sciabstract}

\section{Introduction}
The conformations of low-dimensional physical systems, ranging from polymers to elastic sheets to deposition fronts of crystalline or amorphous phases, are often mathematically described by a corresponding smooth geometry at a continuum level. For instance, a one-dimensional (1D) space curve is used to represent a polymer undergoing Brownian dynamics~\cite{randomwalk},  a two-dimensional (2D) surface models an interface where two bulk phases coexist in three dimensions (3D)~\cite{PFC} or membranes that are embedded in and move through the third dimension \cite{Safran}. These geometric representations are valid when out-of-plane deformations occur on length scales much larger than the thickness of the filament or membrane (and when the curvatures are also relatively small), such as when a cell membrane or a graphene sheet with a thickness $a$ deforms at a wavelength $\lambda$ where $\lambda\gg a$~\cite{Geim1, Geim2, graphene1, graphene2}. 

The physics of such systems is determined not only by their intrinsic dimensionality $d\,,$ but also by the dimension of the embedding domain $D\,.$ For example, in the Frenet-Serret frame of a smooth closed curve~\cite{Stoker, Millman, Kamien}, the dynamics of a ring polymer confined to a plane ($d=1\,,$ $D=2$) requires again only one physical condition for a velocity along the curve normal, whereas the dynamics of a ring polymer in space ($d=1\,,$ $D=3$) would be determined by two velocity components along the curve normal and curve binormal. Similarly, the interface growth dynamics of a crystal in space ($d=2\,,$ $D=3$) is fully determined by one physical condition, i.e. the deposition velocity along the normal direction to the interface~\cite{Brower1}. At equilibrium the physical conditions would be provided by the  Euler-Lagrange equations that minimize the corresponding free energy, while for non-equilibrium systems, we need to replace these by appropriate dynamical equations; either of these must be consistent with the $D-d$ physical relations that are needed to fully specify the state of the closed geometry of a system.

Following the theory introduced in Ref.~\cite{Kaplan} for the controlled growth and form of bioinspired coprecipitation patterns of carbonate and silica~\cite{Ruiz1, Ruiz3, Wim, Kellermeier}, here we develop a geometrical theory for the constrained growth and form of a non-planar smooth surface with material deposition at its curvilinear edges.  Our formulation addresses the deposition of a surface that is laid down by a closed space curve ($d=1\,,$ $D=3$), for which two physical conditions are needed to determine the dynamics and form: The first condition corresponds to the growth velocity along the curve normal (the growth direction) must be specified. For the second condition, surface smoothness demands that a velocity component along the surface normal be prohibited. Instead, the time variation of the curve normal can have a component along the surface normal, which specifies an extrinsic curvature for the curling of the front. It is this curvature that needs to be dictated by a second condition for a smooth surface growing at its edge. These two physical conditions complement the geometric compatibility conditions for a smooth surface, i.e., the Codazzi-Mainardi Equations~\cite{Stoker} and the Gauss {\it theorema egregium}~\cite{Stoker}, which we express in a frame co-moving with the curvilinear front. This leads to equations for the dynamics of a space curve "constrained" to a smooth surface that it leaves behind in its wake. We note that the dynamics thus prescribed is fundamentally different from that of a curve freely evolving in space via the normal and binormal velocities, and with compatibility conditions given by the Frenet equations for space curves~\cite{Stoker, Millman, Kamien}.

The theory of constrained surface growth that we present here is relevant to many natural systems all of which have one dimension that is much smaller than the other two: centimeter- to meter-scale molluscan and brachiopod shells recording local shape changes during accretion~\cite{shells1, shells2, shells3, shells4, shells5}, and chemical precipitates, such as micron- to millimeter-scale carbonate-silica composite walls~\cite{Ruiz1, Ruiz3, Wim, Kellermeier} or chemical gardens of thin-walled millimeter- to meter-scale membraneous tubes including underwater hydrothermal vents~\cite{Gardens1, Gardens2}. Mathematically, these structures exhibit several common properties: Growth is strongly localized along an interface of the emerging high-aspect-ratio wall that can be approximated as a 2D smooth surface, and the resulting structures achieve simply connected yet intricate surface geometries. Under the smooth surface assumption, these unifying characteristics imply universal mathematical constraints (due to geometric compatibility) imposed on the growth dynamics and final form of infinitesimally thin surfaces. To quantify the growth and form of these effectively 2D systems, current theoretical approaches often limit the analysis to prescribed geometries or single-valued surface height functions~\cite{shells2, shells5, Steinbock1, Steinbock2}. Here, by deploying a self-consistent covariant theory, we provide a geometrical theory that is capable of describing a range of precipitating patterned structures and capturing their complex, absolute-scale-free morphologies.

 In Section~\ref{sec:theory}, we detail our theory by first formulating the geometry and dynamics of a curvilinear front that leaves behind a smooth surface. In this section, we further derive the geometric compatibility equations in Section~\ref{sec:theory}~\ref{sec:compatibility} in a frame co-moving with the growth front. In Section~\ref{sec:theory}~\ref{sec:closure}, we introduce the mathematical closure relations required for a well-posed problem. In Section~\ref{sec:results}, we show the range of morphologies that result by solving the complete set of equations for the geometrical dynamics of edge-driven surface growth. In Section~\ref{sec:discussion} we conclude with a brief discussion and potential future directions.

\section{Theory of edge-driven growth of a smooth surface}
\label{sec:theory}

\begin{figure}[!ht]
\centering
\includegraphics[width=1\textwidth, clip=true]{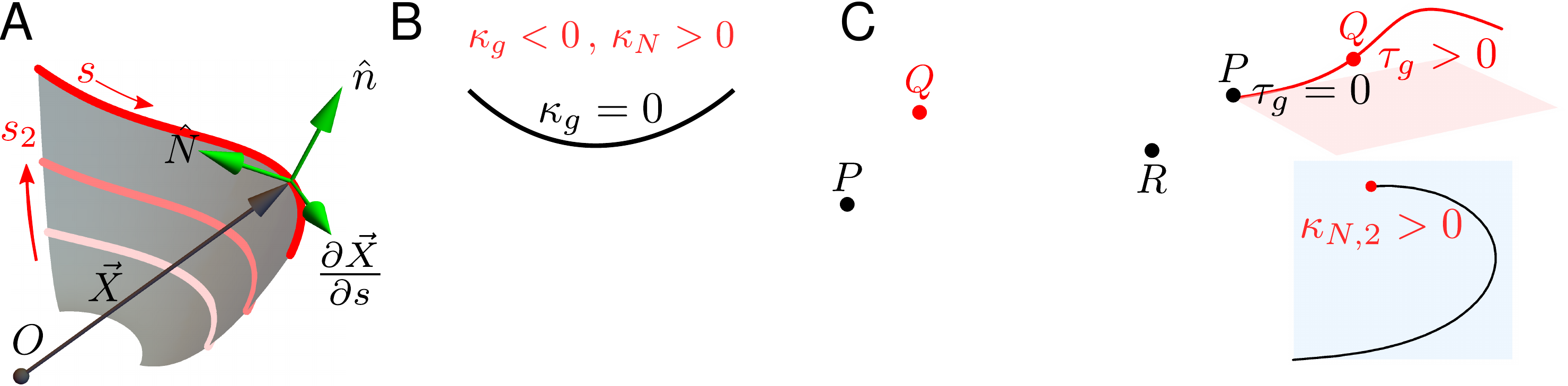}
\caption{{\bf Growth of a surface.} (A) The growth front of a thin wall is represented in terms of a space curve (shades of red) with a position vector $\mathbf{\vec{X}}(s, t)\,,$ where $s$ is the position along the curve and $s_2$ is proportional to the time coordinate $t\,.$ The tangent vector of the curve $\partial \mathbf{\vec{X}}/\partial s\,,$ the surface normal $\mathbf{\hat{N}}\,,$ and the growth direction $\mathbf{\hat{n}}$ form an orthonormal triad. (B) The geodesic curvature $\kappa_g$ is the curvature of a line with respect to a geodesic ($\kappa_g=0$) on a surface. The boundary curve acquires a finite normal curvature $\kappa_N$ when e.g. a plane is folded into a cone. (C) In the absence of the twist of the orthonormal triad along the curve, a finite geodesic torsion $\tau_g$ distinguishes a space curve (e.g. at point $Q$) from a section of a plane curve with $\tau_g=0$ (e.g. at point P on the light red plane). The second normal curvature $\kappa_{N, 2}$ characterizes the curling of the surface at the front (black curve on the light blue plane).}
\label{fig:schematics}
\end{figure}

Our theory considers the growth and form of a two dimensional (2D) non-planar smooth surface in three dimensional (3D) Euclidean space driven by localized growth along a curvilinear edge that is the active boundary of the surface. Then the temporal wake (history) of the curve (Fig.~\ref{fig:schematics}~A, shades of red) constitutes the surface as it is laid out in time (Fig.~\ref{fig:schematics}~A, grey). Defining $t$ as the time variable, $U$ as the local Lagrangian growth speed, and $\hat{\mathbf{n}}$ as the growth direction, the equation of motion for the position vector field of the boundary curve $\vec{\mathbf{X}}$ is given by (Fig.~\ref{fig:schematics}~A)
\begin{equation}
    \label{eq:Sec11}
    \frac{d \vec{\mathbf{X}}}{d t}=\hat{\mathbf{n}}U\,.
\end{equation}
In Eq.~\ref{eq:Sec11}, surface smoothness demands that the growth direction $\hat{\mathbf{n}}$ is a tangent vector to the surface. This eliminates a velocity component parallel to the local surface normal $\mathbf{\hat{N}}\,,$ defined in Fig.~\ref{fig:schematics}~A. Furthermore, although there can be a velocity component along the tangent of the curve $\partial \vec{\mathbf{X}}/\partial s$ ($s:$ arc length coordinate along the curve), this does not change the shape of the curve or the surface, so we define the growth direction $\hat{\mathbf{n}}$ to be orthogonal to both $\partial \vec{\mathbf{X}}/\partial s$ and $\hat{\mathbf{N}}$ (Fig.~\ref{fig:schematics}~A). Equivalently, dropping the tangential growth component reflects a gauge invariance for closed curves ~\cite{Brower1}, which is the limit that we employ here. We note that by adding a tangent speed $V$ along $\partial \vec{\mathbf{X}}/\partial s$ to the right-hand side of Eq.~\ref{eq:Sec11}, i.e. of the form $\hat{\mathbf{n}}U+ (\partial \vec{\mathbf{X}}/\partial s) V\,,$ it is straightforward to generalize our framework to the motion of open curves that leave behind smooth surfaces.

To derive the self-consistent equations of motion governing the spatial configuration and temporal evolution of the boundary curve from Eq.~\ref{eq:Sec11}, we need to define the differential geometric quantities that determine the dynamic configuration of the curve in space and time. To that end, we first define $u^i\equiv\{\sigma, t|i=1\,,2\}$ where $\sigma$ is a fixed parametrization along the curve and $t$ is the time variable. Then, $\vec{\mathbf{X}}=\vec{\mathbf{X}}(\sigma, t)\,,$ and $\vec{\mathbf{X}}_{i}\equiv \partial_{u^i}\vec{\mathbf{X}}\,,$ $\vec{\mathbf{X}}_{ik}\equiv \partial_{u^i}\partial_{u^k}\vec{\mathbf{X}}$ are defined as first and second derivatives of $\vec{\mathbf{X}}\,,$ respectively, where $\partial/\partial t= d/d t$ due to the parametrization of $\vec{\mathbf{X}}(\sigma, t)\,.$ In the following, all letter indices take the values $1, 2$ corresponding to the coordinates $\sigma$ and $t\,,$ respectively, and Einstein summation convention (sum over repeated indices) is used. Importantly, because $\sigma$ is time independent, the mixed partial derivatives satisfy the equality
\begin{equation}
\label{eq:CMeq00}
\frac{\partial}{\partial\sigma}\frac{\partial }{\partial t}=\frac{\partial}{\partial t}\frac{\partial}{\partial\sigma}\,.
\end{equation} 
Next, we introduce the differential geometric variables to evaluate the temporal dynamics of a growing non-planar surface at its curve front. The elements of the $2\times2$ metric tensor (first fundamental form) and their inverse are defined as
\begin{equation}\label{CMeq0a}
g_{ij}\equiv \vec{\mathbf{X}}_{i}\cdot \vec{\mathbf{X}}_{j}\,,\quad g^{ij}\equiv \left(g_{ij}\right)^{-1}\,,
\end{equation}

The second derivatives $\vec{\mathbf{X}}_{ik}$ necessitate the introduction of the Christoffel symbols $\Gamma_{ik}^l$ and the coefficients of the second fundamental form $L_{ik}$ that are defined by
\begin{equation} \label{CMeq0b}
\vec{\mathbf{X}}_{ik}\equiv\Gamma_{ik}^l\vec{\mathbf{X}}_l+L_{ik}\hat{\mathbf{N}}\,. 
\end{equation}
Eq.~\ref{CMeq0b} then yields
\begin{linenomath}\begin{equation}
\Gamma^l_{ik}=\vec{\mathbf{X}}_{ik}\cdot\vec{\mathbf{X}}_m g^{ml}\,,\quad L_{ik}=\vec{\mathbf{X}}_{ik}\cdot\hat{\mathbf{N}}\,.    
\end{equation}\end{linenomath}
Evaluating Eqs.~\ref{CMeq0b}, each of the Christoffel symbols $\Gamma_{ik}^l$ and the coefficients of the second fundamental form $L_{ik}$ are expressed in terms of the 6 dependent scalar variables (see Figs.~\ref{fig:schematics} B, C, and Table~\ref{table:geometry}): the metric of the curve $\sqrt{g}\,,$ the geodesic curvature $\kappa_g\,,$ the normal curvature $\kappa_N\,,$ the geodesic torsion $\tau_g\,,$ the second normal curvature $\kappa_{N,2}\,,$ and the growth speed $U\,:$
\begin{align}
\begin{split}
    \label{CMeq2a}
    &\Gamma_{11}^1=\frac{\partial \sqrt{g}}{\partial s}\,,\quad \Gamma_{11}^2=\frac{g}{U}\kappa_g\,,\quad \Gamma_{22}^1=-\frac{U}{\sqrt{g}}\frac{\partial U}{\partial s}\,,\quad \Gamma_{22}^2=\frac{1}{U}\frac{\partial U}{\partial t}\,,\\
    &\Gamma_{12}^1=\Gamma_{21}^1=-U\kappa_g\,,\quad \Gamma_{12}^2=\Gamma_{21}^2=\frac{\sqrt{g}}{U}\frac{\partial U}{\partial s}\,,
\end{split}
\end{align}

\begin{align}
\begin{split}\label{CMeq2b}
    L_{11}=g\kappa_N\,,\quad L_{22}=U^2\kappa_{N,2}\,,\quad L_{12}=L_{21}=\sqrt{g}U\tau_g\,.
\end{split}    
\end{align}

\subsection{Geometric compatibility equations}
\label{sec:compatibility}

The fundamental theorem of surface geometry \cite{Stoker, Millman} demands that a necessary condition for the first and second fundamental forms to be consistent with a surface is the satisfaction of  geometric compatibility conditions that relate $\Gamma^l_{ik}\,,$ $L_{ij}\,,$ and their derivatives, thereby ensuring the existence of a smooth surface (with continuous third derivatives of $\vec{\mathbf{X}}$) in 3-space. These are the well-known Codazzi-Mainardi and Gauss equations, which as we will see lead to 3 independent dynamical equations for 3 of the dependent scalar variables, specifically, for $\kappa_g\,,$ $\kappa_N\,,$ and $\tau_g\,.$ We will separately determine a dynamical equation of motion for $\sqrt{g}$  (see Eq.~\ref{CMeq9} and Table~\ref{table:simulations}). This leaves us requiring two more equations for closure that need input from physical chemistry. These two relations, for the second-normal curvature $\kappa_{N,2}\,,$ and the edge-curve velocity $U\,,$ are motivated by a combination of experimental growth mechanisms as well as symmetry relations. All together, this leads to a total of 6 equations for the 6 dependent scalar variables $\sqrt{g}(s,t),\kappa_g (s,t), \kappa_N (s,t), \tau_g(s,t), \kappa_{N,2}(s,t), U(s,t)$ and constitute a self-consistent model for the form of smooth surfaces driven by growth along their free edge.

\subsubsection{Curve metric evolution in a moving frame}

To derive an equation of motion for the metric of the curve $\sqrt{g}$, we take the time derivative of $g\equiv g_{11}=\vec{\mathbf{X}}_1\cdot \vec{\mathbf{X}}_1\,.$ Using Eq.~\ref{eq:CMeq00}, taking the derivative of Eq.~\ref{eq:Sec11} with respect to $\sigma\,,$ and substituting the result in $\partial g/\partial t$ yields
\begin{equation}
\label{CMeq7}
\frac{\partial g}{\partial t}=2 \frac{\partial \vec{\mathbf{X}}}{\partial \sigma}\cdot \frac{\partial}{\partial \sigma}\left(\mathbf{\hat{n}}U\right)\,.
\end{equation}
Then, rewriting the spatial derivatives in terms of $\partial/\partial s$, which satisfies $\sqrt{g}\partial/\partial s=\partial/\partial\sigma\,,$ and the orthogonality relation $\mathbf{\hat{n}}\perp \partial\vec{\mathbf{X}}/\partial s$ transform Eq.~\ref{CMeq7} to
\begin{equation}
\label{CMeq8}
\frac{d g}{d t}=2 g U \frac{\partial \vec{\mathbf{X}}}{\partial s}\cdot \frac{\partial\mathbf{\hat{n}}}{\partial s}=-2gU \frac{\partial^2 \vec{\mathbf{X}}}{\partial s^2}\cdot\mathbf{\hat{n}}\,.
\end{equation}
In Eq.~\ref{CMeq8} we have first used the product rule $\vec{\mathbf{a}}\cdot\vec{\mathbf{b}}'=(\vec{\mathbf{a}}\cdot\vec{\mathbf{b}})'-\vec{\mathbf{a}}'\cdot\vec{\mathbf{b}}$ for two arbitrary vectors $\vec{\mathbf{a}}\,,$  $\vec{\mathbf{b}}$ (where accents denote derivatives). Then, to obtain the second equality, we used $\mathbf{\hat{n}}\perp \partial\mathbf{\vec{X}}/\partial s$ since for any unit vector $\hat{\mathbf{a}}\,,$ 
\begin{equation}
\label{eq:SI7}
    \hat{\mathbf{a}}\cdot\hat{\mathbf{a}}'=0
\end{equation}
always holds. Substituting $\kappa_g\equiv \hat{\mathbf{n}}\cdot\partial^2\vec{\mathbf{X}}/\partial s^2\,,$ Eq.~\ref{CMeq8} becomes
\begin{equation}
\label{CMeq9}
\frac{\partial \sqrt{g}}{\partial t}=-\sqrt{g}\kappa_g U\,.
\end{equation}
Eq.~\ref{CMeq9} governs the time dependence of the metric, in other words, the change of the local arc length during growth. 

\subsubsection{Codazzi-Mainardi equations in a moving frame}
\label{sec:Codazzi}

Two of the three compatibility equations are given in the closed form as~\cite{Stoker}
\begin{equation}
    \label{CMeq1}
\frac{\partial L_{ik}}{\partial u^j}-\frac{\partial L_{ij}}{\partial u^k}+\Gamma^l_{ik} L_{ij}-\Gamma^l_{ij} L_{ik}=0\,.
\end{equation}
Eq.~\ref{CMeq1} is trivially satisfied when $j=k\,.$ Then, taking $j=1\,, k=2$ ($j=2\,, k=1$ multiplies Eq.~\ref{CMeq1} by -1), and setting $i=1\,,2$ successively, returns the two compatibility equations, known as the Codazzi-Mainardi equations~\cite{Stoker}:

\begin{align} \label{CMeq2}
\begin{split}
& \frac{\partial L_{12}}{\partial \sigma}-\frac{\partial L_{11}}{\partial t}+\Gamma^1_{12} L_{11}+\Gamma^2_{12} L_{21}-\Gamma^1_{11} L_{12}-\Gamma^2_{11} L_{22}=0\,, \\
&\frac{\partial L_{22}}{\partial \sigma}-\frac{\partial L_{21}}{\partial t}+\Gamma^1_{22} L_{11}+\Gamma^2_{22} L_{21}-\Gamma^1_{21} L_{12}-\Gamma^2_{21} L_{22}=0\,.
\end{split} 
\end{align}

\begin{table*}[ht!]
\begin{center}
\small{\begin{tabular}{|c||c|c|}
 \hline
   Variable & Definition & Description \\
   \hline\hline
   $\sigma$ & $\sigma\in \left[0, \sigma_{max}\right]$ & fixed coordinate along\\
   & & the boundary curve\\ \hline
   $t$ &  $t\geq0$ & time\\ \hline
   $g\equiv g_{11}$ & $\left(\partial\vec{\mathbf{X}}/\partial \sigma\right)^2$ & metric of the boundary curve\\ \hline
   $g\equiv g_{22}$ & $\left(\partial\vec{\mathbf{X}}/\partial t\right)^2$ & metric of the orthogonal curve along $\hat{\mathbf{n}}$\\ \hline
   $U$ &  & growth speed\\ \hline
   $g_{ij}$ &  
   $\begin{bmatrix}
   g & 0 \\
   0 & U^2
   \end{bmatrix}$
   & metric tensor\\ \hline
   $g^{ij}$ &  
   $\begin{bmatrix}
   1/g & 0 \\
   0 & 1/U^2
   \end{bmatrix}$
   & inverse of the metric tensor\\ \hline
   $ds$ & $\sqrt{g}d\sigma$ & local arc length \\ \hline
   $\partial\vec{\mathbf{X}}/\partial s$ & $\sqrt{g}^{-1} \partial\vec{\mathbf{X}}/\partial \sigma$ & unit vector along the boundary curve (Fig.~\ref{fig:schematics}~A)\\ \hline
   $\hat{\mathbf{n}}$& $\hat{\mathbf{n}}= U^{-1}\left(\partial\vec{\mathbf{X}}/\partial t \right)$ & unit vector along the growth direction (Fig.~\ref{fig:schematics}~A)\\ \hline
   $\kappa_g$ & $\mathbf{\hat{n}}\cdot \partial^2 \vec{\mathbf{X}}/\partial s^2$ & geodesic curvature (Fig.~\ref{fig:schematics}~B)\\ \hline
   $\kappa_N$ & $\mathbf{\hat{N}}\cdot \partial^2 \vec{\mathbf{X}}/\partial s^2$ & normal curvature (Fig.~\ref{fig:schematics}~B)\\ \hline
   $\tau_g$ & $\mathbf{\hat{N}}\cdot \partial \mathbf{\hat{n}}/\partial s$ & geodesic torsion (Fig.~\ref{fig:schematics}~C) \\ \hline
   $\kappa_{N, 2}$ & $\mathbf{\hat{N}}\cdot  \partial\mathbf{\hat{n}}/U\partial t$ & second normal curvature (Fig.~\ref{fig:schematics}~C) \\ \hline
\end{tabular}}
\end{center}
\caption{{\bf Definitions of the scalar and vector geometric variables}. The independent variables $\sigma\,, t$ and the remaining dependent variables listed here determine the configuration of the boundary curve in space and time.}
\label{table:geometry}
\end{table*}

\noindent
Substituting Eqs.~\ref{CMeq2a} and~\ref{CMeq2b} into Eqs.~\ref{CMeq2} returns the Codazzi-Mainardi equations in a frame co-moving with the front at a speed $U\,:$
\begin{equation}\label{CMeq2c}
    \frac{\partial \kappa_N}{\partial t}=\frac{\partial}{\partial s}\left(U\tau_g\right)+\tau_g\frac{\partial U}{\partial s}+\kappa_g U\left(\kappa_N-\kappa_{N,2}\right)\,,
\end{equation}
    
\begin{equation}\label{CMeq2d}
    \frac{\partial \tau_g}{\partial t}=\frac{\partial}{\partial s}\left(U\kappa_{N,2}\right)-\frac{\partial U}{\partial s}\kappa_N+2U\kappa_g\tau_g\,.
\end{equation}
Eqs.~\ref{CMeq2c} and~\ref{CMeq2d} are the two of the three compatibility equations that govern the spatial and temporal configuration of the growth front represented by a space curve embedded in a growing smooth surface.

\subsubsection{The Gauss {\it theorema egregium} in a moving frame}
\label{sec:GaussBonnet}

\begin{figure}[!ht]
\centering
\includegraphics[width=0.5\textwidth, clip=true]{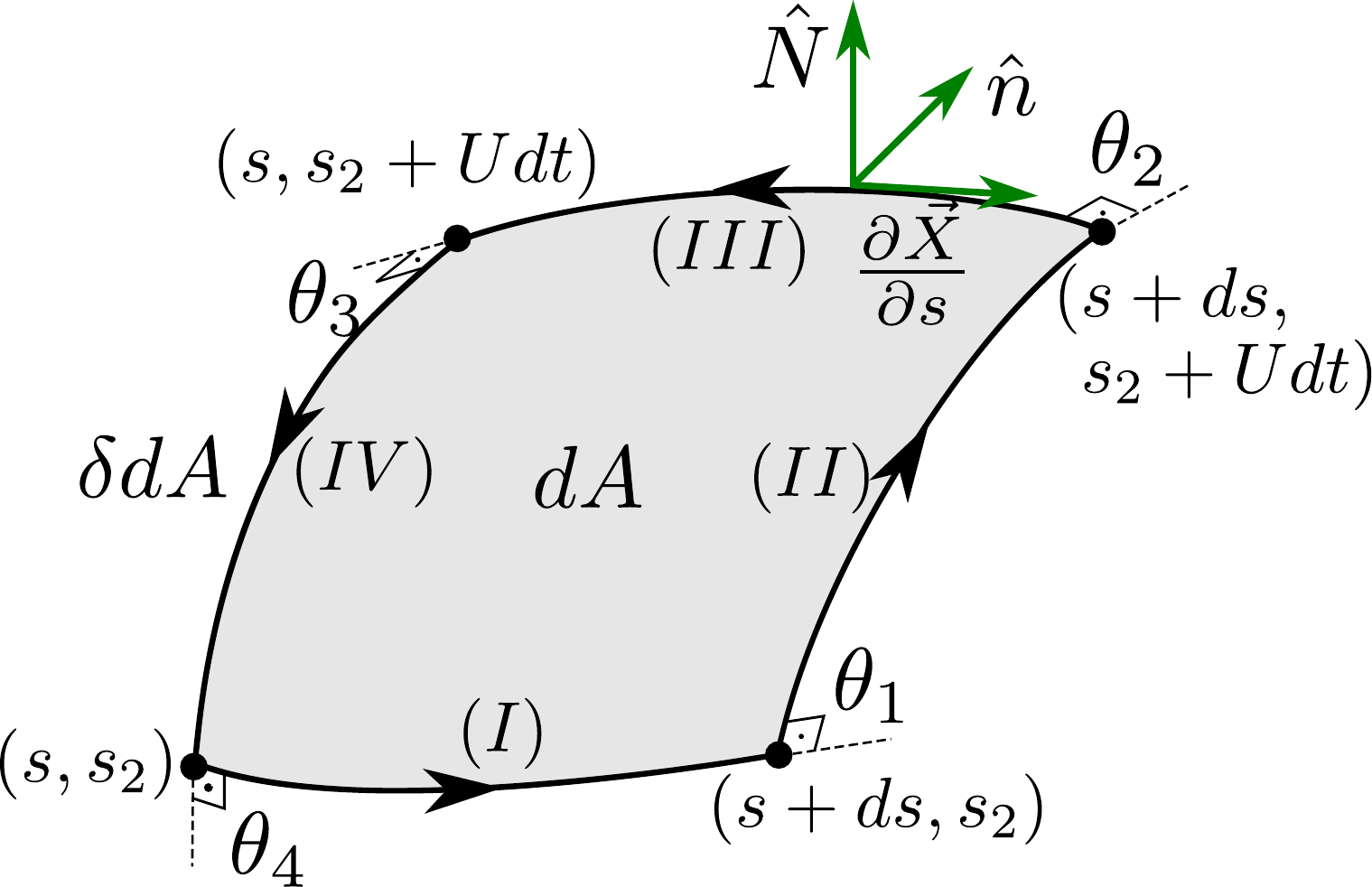}
\caption{{\bf Gauss-Bonnet theorem on an infinitesimal surface patch dA.} The boundary $\delta \Omega\equiv\delta dA$ consists of four curvilinear sections, labeled from $(I)$ to $(IV)\,,$ and is parametrized by two coordinates $s_1 \equiv s$ and $s_2\sim t\,.$ The orthonormal triad is shown by green arrows. The positive angles are defined in the counter-clockwise direction along the integration path, and $\theta_1\,,$ $\theta_2\,,$ $\theta_3\,,$ $\theta_4$ are right angles.}
\label{fig:schematics2}
\end{figure}

The third compatibility equation for the existence of a smooth surface patch $\Omega$ relates its Gaussian curvature $\kappa_G$ to the geodesic curvature of its boundary $\delta \Omega\,,$ $\tilde{\kappa}_g.$ This relation is known in integral form as the Gauss-Bonnet theorem, which, for a simply connected surface patch $\Omega$ is given as ($\theta_i:$ angles at the vertices along the boundary $\delta\Omega\,;$ $\theta_1=\theta_2=\theta_3=\theta_4=\pi/2\,;$ see Fig.~\ref{fig:schematics2})
\begin{equation}
    \label{CMeq3}
    \int_\Omega \kappa_G dA-\oint_{\delta\Omega} \tilde{\kappa}_g d\tilde{s}+\sum_{i=1}^4 \theta_i=2\pi\,.
\end{equation}

The differential formulation of Eq.~\ref{CMeq3} can be derived by defining $\sigma_1\equiv \sigma\,,$ $\sigma_2\equiv t\,.$ The infinitesimal arc lengths along the coordinates $\sigma\,, t$ then become $ds_1\equiv ds=\sqrt{g}d\sigma$ and $ds_2\equiv U dt\,,$ respectively, where, $g\equiv g_{11}$ and $U^2=g_{22}$ from Eq.~\ref{CMeq0a} (see Table~\ref{table:geometry}). About an infinitesimal area element $dA\,,$ traversing its boundary $\delta dA$ in the counter-clockwise direction (Fig.~\ref{fig:schematics2}), the line elements of the four curve sections along $\delta dA$ are given as
\begin{equation}
\label{CMeq3a}
d\tilde{s}_{(I)}=ds\,,\quad
d\tilde{s}_{(II)}=Udt\,,\quad
d\tilde{s}_{(III)}=-ds\,,\quad
d\tilde{s}_{(IV)}=-Udt\,,\quad    
\end{equation}
Then, the infintesimal area element is found as $dA=\sqrt{g}Ud\sigma dt\,,$ and the geodesic curvatures of each of the four curve sections become

\begin{equation}
\tilde{\kappa}_{g}^{(I)}=-\kappa_g\,, \quad \tilde{\kappa}_{g}^{(II)}=-\frac{1}{U}\frac{\partial U}{\partial s}\,, \quad
\tilde{\kappa}_{g}^{(III)}=-\kappa_g\,, \quad
\tilde{\kappa}_{g}^{(IV)}=-\frac{1}{U}\frac{\partial U}{\partial s}\,.    
\label{CMeq3b}
\end{equation}
By using Eqs.~\ref{CMeq3a} and~\ref{CMeq3b}, the integral of $\tilde{\kappa}_g$ over $\delta dA$ (the second term on the left-hand side of Eq.~\ref{CMeq3}) is rewritten as
\begin{linenomath}\begin{align}
\begin{split}
    \oint_{\delta dA}\tilde{\kappa}_g d\tilde{s}&=-\int_{(I)}(\kappa_g\sqrt{g})\bigg|_{s_1, s_2} d\sigma'-\int_{(II)}\frac{\partial U}{\partial s}\bigg|_{s_1+ds_1, s_2}dt'\\&+\int_{(III)}(\kappa_g\sqrt{g})\bigg|_{s_1, s_2+ds_2} d\sigma'+\int_{(IV)}\frac{\partial U}{\partial s}\bigg|_{s_1, s_2}dt'\\
    &=\int_{s_1}^{s_1+\sqrt{g}d\sigma} \left[(\kappa_g\sqrt{g})\bigg|_{s_1, s_2+ds_2}-(\kappa_g\sqrt{g})\bigg|_{s_1, s_2}\right]d\sigma'\\&+\int_{s_2}^{s_2+U dt}\left[\frac{\partial U}{\partial s}\bigg|_{s_1, s_2}-\frac{\partial U}{\partial s}\bigg|_{s_1+ds_1, s_2}\right]dt'\,,
\end{split}
\label{CMeq4}
\end{align}\end{linenomath}
where the accents are used to distinguish between the integration constants and the boundaries of integration. Reformulating the expressions inside the square brackets of Eq.~\ref{CMeq4} in terms of definite integrals
\begin{linenomath}\begin{align}
\label{CMeq5}
\begin{split}
&\left(\kappa_g\sqrt{g}\right)(s_1, s_2+Udt)-\left(\kappa_g\sqrt{g}\right)(s_1, s_2)=\int^{s_2+U dt}_{s_2} U dt'\frac{1}{U}\frac{\partial }{\partial t'}\left(\kappa_g\sqrt{g}\right)\,,\\
&\frac{\partial U}{\partial s}(s_1+\sqrt{g}d\sigma, s_2)-  \frac{\partial U}{\partial s}(s_1, s_2)=\int^{s_1+\sqrt{g}d\sigma}_{s_1}\sqrt{g}d\sigma' \frac{\partial^2 U}{\partial s'^2}\,,  
\end{split}
\end{align}\end{linenomath}
and dropping the accents, Eqs.~\ref{CMeq4} and~\ref{CMeq5} together yield
\begin{equation}
\label{CMeq6}
      \oint_{\delta dA}\tilde{\kappa}_g d\tilde{s}=\int_{dA}  \left[\frac{1}{U}\frac{\partial \kappa_g}{\partial t}\sqrt{g}+\frac{1}{U}\kappa_g\frac{\partial \sqrt{g} }{\partial t}-\frac{1}{U}\frac{\partial^2 U}{\partial s^2}\sqrt{g}\right]U d\sigma dt\,.
\end{equation}
For an arbitrary surface patch $\Omega \gg dA\,,$ Eq.~\ref{CMeq3},~\ref{CMeq6},~\ref{CMeq9} and $\sum_{i=1}^4 \theta_i=2\pi$ (see Fig.~\ref{fig:schematics2}) together result in

\begin{equation}
\label{CMeq10}
\int_{\Omega}  \left[\kappa_G-\frac{1}{U}\frac{\partial \kappa_g}{\partial t}+\kappa_g^2+\frac{1}{U}\frac{\partial^2 U}{\partial s^2}\right]\sqrt{g}Ud\sigma dt=0\,.
\end{equation}
In order for the integral in Eq.~\ref{CMeq10} to vanish on every infinitesimal surface patch $dA$ of the finite surface $\Omega$, the integrand must be equal to zero, that is,
\begin{equation}
\label{CMeq11}
\frac{\partial \kappa_g}{\partial t}=\frac{\partial^2 U}{\partial s^2}+\left(\kappa_g^2+\kappa_G\right) U\,.
\end{equation}
Eq.\ref{CMeq11} is the Gauss {\it theorema egregium}, i.e., differential formulation of the Gauss-Bonnet theorem, for a simply connected smooth surface in a frame co-moving with the front.

Eqs.~\ref{CMeq2c},~\ref{CMeq2d}, and~\ref{CMeq11} constitute the three geometric compatibility equations that must hold at the growth front of a surface to maintain its smoothness. They govern, respectively, the dynamics of the extrinsic scalar variables $\kappa_N\,,$ $\tau_g$ that depend on the local surface orientation $\hat{\mathbf{N}}\,,$ and the intrinsic scalar variable $\kappa_g$ that is independent of $\hat{\mathbf{N}}$ (see Table~\ref{table:geometry}). That way, the instantaneous configuration of the growth site is coupled to the extrinsic geometry of the embedding surface determined by $\hat{\mathbf{N}}\,,$ as well as the intrinsic geometries of the surface and the curve itself that are independent of $\hat{\mathbf{N}}\,.$ To complete the formulation of the problem, Eqs.~\ref{CMeq9},~\ref{CMeq2c},~\ref{CMeq2d}, and~\ref{CMeq11} need to be complemented with two closure relations that set the extrinsic growth speed $U$ and the second normal curvature $\kappa_{N,2}\,,$ which is also an extrinsic variable.

\subsection{Closure relations}
\label{sec:closure}

\subsubsection{Constitutive equation for edge-curve speed}
\label{sec:speed}

We determine the local growth speed $U$ by a power series expansion in terms of $\kappa_g\,,$ its second derivative $\partial^2 \kappa_g/\partial s^2\,,$ $\kappa_N\,,$ $\tau_g\,,$ and $\kappa_{N, 2}\,.$ The primary reason for the power series approximation is that it simplifies the analysis by assuming growth dynamics localized to the boundary curve. Furthermore, there is a physical motivation for this approximation: it reproduces qualitatively the dynamics of diffusion-limited growth, in analogy with the geometrical models of dendritic solidification~\cite{Brower1}. Because dendritic growth amplifies local perturbations along the boundary curve, the surfaces resulting from our theory can form highly intricate shapes. For reasons clarified below, we choose to truncate the expansion at the third order in an inverse characteristic length scale $1/\ell\,,$ where $\ell$ sets the linear dimensions of the initial condition of the structures in the simulations. In fact, the only length scale in the problem is $\ell\,,$ therefore, we set $\ell=1$ in dimensionless units. The series expansion of the growth speed up to third order in $1/\ell$ (in dimensionless units) is given by
\begin{eqnarray}
\label{eq:Sec21}
\nonumber
&U=&-\alpha_1 \kappa_g+\alpha_2 \kappa_g^2+\alpha_3 \kappa_g^3+\eta_1 H^2+\eta_{21} \kappa_N \kappa_{N,2}+\eta_{22}\tau_g^2+\eta_3\kappa_g H^2 \\ & & -\eta_{41}\kappa_g \kappa_N \kappa_{N,2}+\eta_{42}\kappa_g\tau_g^2+\eta_5\tau_g H+\eta_6\kappa_g\tau_gH-\lambda \frac{\partial^2 \kappa_g}{\partial s^2}+\mathcal O (\ell^{-4})\,,
\end{eqnarray} 
where $H\equiv (\kappa_N+\kappa_{N, 2})/2$ is the mean curvature of the surface, and the coefficients of each term are positive scalars. The term $\partial^2 \kappa_g/\partial s^2$ suppresses unstable outward kinks along the boundary curve that originate due to the terms with $\kappa_g\,,$ analogous to the Mullins-Sekerka instability in dendritic solidification~\cite{MullinsSekerka}. Thus, the prefactor $\lambda$ multiplying $\partial^2 \kappa_g/\partial s^2$ is proportional to the line tension along the growth front~\cite{Brower1}. 

The power series given in Eq.~\ref{eq:Sec21} implies that growth may continue indefinitely albeit with an ever decreasing speed e.g. while a vase with a uniform circular boundary grows, where only $\kappa_g$ and $\kappa_N$ are finite (Fig.~\ref{fig:schematics}~B)~\cite{Kaplan}. In Eq.~\ref{eq:Sec21}, $\kappa_g$ breaks the $\mathbf{\hat{n}}\rightarrow -\mathbf{\hat{n}}$ symmetry and is present at all orders, i.e. $-\mathbf{\hat{n}}$ points into the surface already laid down, whereas $\mathbf{\hat{n}}$ is the direction of growth. Additionally, there are no first order derivatives in the arc length coordinate $s$ because $U$ must remain unchanged under the transformation $s\rightarrow-s\,.$ In contrast with the geodesic curvature $\kappa_g\,,$ the extrinsic geometrical quantities $\kappa_N\,,$ $\tau_g\,,$ and $\kappa_{N, 2}$ appear as even terms since they change sign under $\mathbf{\hat{N}}\rightarrow - \mathbf{\hat{N}}$ transformation, under which $U$ must remain invariant. In Eq.~\ref{eq:Sec21}, the terms at order $\mathcal O (\ell^{-2})$ increase the growth speed when the associated deformations emerge. When the absolute values of any of $\kappa_N\,,$ $\tau_g\,,$ and $\kappa_{N, 2}$ become very large, then the terms penalizing the speed can enter the expansion at the third order, namely by the product of any pair of these variables with $\kappa_g\,.$ Thus, all of the third order terms in Eq.~\ref{eq:Sec21} are mainly responsible for decreasing the growth rate, including the terms proportional to $\kappa_g^3$ and $\partial^2 \kappa_g/\partial s^2\,.$ Exceptions to these may occur when for instance $\eta_3\kappa_g H^2$ reinforces the speed for $\kappa_g>0\,$; however it is in general balanced by the other third order terms for a variety of sculptures simulated in this work. An expansion including the fourth order terms in $1/\ell$ could shift the speed penalty terms to this order, but such an expansion would produce many additional free parameters, making the numerical implementation and analysis tedious, without additional insight.

\subsubsection{Constitutive equation for surface second-normal curvature}
\label{sec:2ndcurvature}

The second normal curvature $\kappa_{N, 2}$ is associated with local curling along the growth direction $\hat{\mathbf{n}}\,,$ i.e., it is the extrinsic curvature of any curve locally parallel to $\hat{\mathbf{n}}$ on the surface (Fig.~\ref{fig:schematics}~C). Intuitively, its evolution must vanish when $U=0\,.$ A simple closure relation that satisfies this requirement is  a time evolution equation for the mean curvature $H\,,$ given as
\begin{equation}
\frac{\partial H}{\partial t}=\zeta U H\,,\quad H\equiv \frac{1}{2}\left(\kappa_{N}+\kappa_{N,2}\right)\,.
\label{eq:Sec22}    
\end{equation}
In real units, $\zeta$ would have dimensions of $1/\text{length}\,,$ i.e., it provides a curvature scale over which the growth in $H$ occurs. Eq.~\ref{eq:Sec22} is convenient because, for $\zeta>0\,,$ it can locally amplify any emerging non-uniformity in $\kappa_{N, 2},$ giving rise to reinforced out-of-plane wrinkles of the surface along the boundary curve.  For $\zeta<0$ it evolves towards $H=0$ where the surface becomes locally a minimal surface ($H=0$). Although one can propose an infinite number of closure relations for $\kappa_{N,2}$ that would take into account this dynamics, Eq.~\ref{eq:Sec22} is a very simple one in that it obeys the limit $\partial \kappa_{N,2}/\partial t \rightarrow 0$ when $U\rightarrow 0$ (Eq.~\ref{CMeq2c} already ensures $\partial \kappa_N/\partial t \rightarrow 0$ when $U\rightarrow 0$), it is first order in both $\kappa_N\,, \kappa_{N, 2}$ (ignoring the curvature dependence of $U$) and depends only on a single control parameter $\zeta\,.$

Eqs.~\ref{CMeq9},~\ref{CMeq2c},~\ref{CMeq2d},~\ref{CMeq11},~\ref{eq:Sec21}, and~\ref{eq:Sec22} constitute a mathematical framework for the edge-driven surface growth when complemented by  boundary conditions and initial conditions that are summarized in Table~\ref{table:simulations} and to be discussed in the next section.  Our theory models the dynamics of the curvilinear growth front in a fundamentally different way from that of conventional curves, such as vortex filaments in fluids~\cite{Segur, Hashimoto, KdV}, where only the intrinsic geometry of the curve is relevant. This is because here the configuration of the growth site is coupled with both the extrinsic and intrinsic geometries of the non-planar embedding surface at the curvilinear growth front. 

Previously~\cite{Kaplan}, we utilized a very similar mathematical framework but with a different closure relation instead of Eq.~\ref{eq:Sec22} for the accretionary growth and form of thin-walled composites emerging from BaCO$_3-$SiO$_2$ coprecipitation in basic aqueous solutions. The experimental data for the growth of these composites exhibited $U\sim t^{-1/2}$ and a growth instability with increasing $\kappa_g\,,$ both (i) and (ii) being characteristics of  diffusion-limited growth~\cite{Ruiz1, Kaplan}, thus making Eq.~\ref{eq:Sec21} useful. However, our closure relation for $\kappa_{N,2}$ was taken to be~\cite{Kaplan} 
\begin{equation}
\frac{\partial \kappa_{N,2}}{\partial t}=\gamma \frac{\partial^2\kappa_{N,2}}{\partial s^2}+\tilde{\zeta}\kappa_g\kappa_N U(\kappa_{N,2}-q_b)\,,    
\label{eq:Sec23}  
\end{equation}
where the first term on the right relaxes the curling mode along the growth front with a diffusivity $\gamma\,.$ With a constant $\tilde{\zeta}$ (dimensions: length)\,, the second term (the source term) induces curling due to a coarse-grained bending parameter $q_b\,,$ which was assumed to be inversely proportional to the local $pH$ of the solution. Based on the closure relations~\ref{eq:Sec23} for $\kappa_{N, 2}$ and Eq.~\ref{eq:Sec21} for $U$, which are both experimentally motivated, the geometric compatibility relations Eqs.~\ref{CMeq2c},~\ref{CMeq2d},~\ref{CMeq11}, and the equation governing the curve metric (Eq.~\ref{CMeq9}), the geometrical theory of accretionary growth explained a range of observed morphologies, specifically vaselike, coral-like, and helical precipitates~\cite{Kaplan}. It further predicted pH-dependent sequential growth pathways for new shapes that we then synthesized and employed to build optical waveguides owing to the optical properties of BaCO$_3$ and SiO$_2\,.$ 

In this paper, we have chosen to simplify the closure relation for the second-normal curvature and use Eq.~\ref{eq:Sec22} instead of Eq.~\ref{eq:Sec23}. The two equations differ in three ways, First, a natural curvature scale for curling in Eq.~\ref{eq:Sec23} arises from $\bar{\zeta}\kappa_g\kappa_N\,,$ which is replaced by a single constant $\zeta$ in Eq.~\ref{eq:Sec22}. The factor $\bar{\zeta}\kappa_g\kappa_N$ was chosen to induce curling inwards at an interface with $\kappa_g<0$ at surfaces such as a vase (i.e., a cone; see Fig.~\ref{fig:schematics}~B) and outwards when $\kappa_g>0$ (e.g., on an inverse cone growing at its narrower opening). The dependence of $\bar{\zeta}\kappa_g\kappa_N$ on $\kappa_N$ ensured a higher curling rate at higher normal curvatures, an effect observed in the growth of helical precipitates~\cite{Kaplan}. Second, by defining a P\'{e}clet number $Pe\equiv U L_c/\gamma$ ($L_c:$ time-dependent length of the edge circumference), Eq.~\ref{eq:Sec23} can describe diffusion-driven curling for low $Pe\,,$ whereas Eq.~\ref{eq:Sec23} is strictly constrained to the limit $Pe \rightarrow \infty$ where curling happens locally. Third, Eq.~\ref{eq:Sec23} imposes an experimentally motivated upper limit for curling set by $q_b\,,$ in contrast, Eq.~\ref{eq:Sec22} allows indefinite localized growth in $H$ manifested by strong undulations in $\kappa_N$ along the boundary curve or in $\kappa_{N,2}$ in the time axis.  

Here, we will focus on the simulation of three classes of hypothetical morphologies that exhibit strong undulations in these two extrinsic curvatures. Our results further highlight the versatility of our geometrical approach and ease of its implementation in a one-dimensional fixed domain spanned by the variable $\sigma$ and in time $t\,.$

\section{Results}
\label{sec:results}

\subsection{Simulation procedure}
\label{sec:simulation}

We simulated the geometrically-constrained growth of scale-free smooth surfaces at their free curvilinear boundary, specifically vase-like patterns (Figs.~\ref{fig:res1} and~\ref{fig:res2}, Movies S1-S6), shell-like patterns (Figs.~\ref{fig:res3} and~\ref{fig:res4}, Movies S7-12), and oscillating stem-like structures (Fig.~\ref{fig:res5}, Movies S13-15). The growth dynamics and the final form of the morphologies are based on the solutions of the scalar geometric variables $\sqrt{g}\,,$ $\kappa_g\,,$ $\kappa_N\,,$ $\kappa_{N,2}\,,$ $\tau_g\,,$ governed by Eqs.~\ref{CMeq2c},~\ref{CMeq2d}, and ~\ref{CMeq11} (the three geometric compatibility equations) and Eqs.~\ref{eq:Sec21},~\ref{eq:Sec22} (the closure relations); see Table~\ref{table:simulations}. These six equations constitute a closed set of nonlinear partial differential equations that is fourth order in the fixed coordinate $\sigma$ and fifth order in time $t\,,$ subject to the four Neumann boundary conditions (Table~\ref{table:simulations}) and five initial conditions for $\sqrt{g}\,,$ $\kappa_g\,,$ $\kappa_N\,,$ $\kappa_{N,2}\,,$ $\tau_g\,,$ $U\,.$ For the dimensionless simulation parameters given in Table~\ref{table:simulations} for each class of the morphologies, we numerically solve the equations in $\sigma$ and $t$ by using the FEniCS finite element package on Python 3.6~\cite{fenics}. The initial conditions are determined by the following mathematical procedure: For vases and oscillating stems, we use the following initial condition for the position vector of the front $\vec{X}(\sigma, t)$ ($z:$ height coordinate, $k:$~wave number) at $t=0\,:$
\begin{linenomath}\begin{gather}
\vec{\mathbf{X}}_i (\sigma, z) \equiv \vec{\mathbf{X}} (\sigma, t=0)= \left\{f(\epsilon, m, z, \sigma)\cos(2\pi \sigma)\,, -f(\epsilon, m, z, \sigma)\sin(2\pi \sigma)\,, z\right\}\,, \label{eq:Sec31}\\
f(\epsilon, m, z, \sigma)\equiv 1+m z+\epsilon m^2 \cos(2\pi k \sigma)\,,\quad \sigma\in\left[0\,, 1/k\right]\,,\nonumber
\end{gather}\end{linenomath}
at a height $z=0\,.$ Here, the initial angle between the wall and the $xy-$plane $\beta$ is equal to $\beta=1/m\,.$ The derivatives of Eq.~\ref{eq:Sec31} with respect to $\sigma$ and $z$ yield the tangent vectors to the surface $\partial \vec{\mathbf{X}}/\partial \sigma$ and $\vec{\mathbf{n}}\equiv\sqrt{1+m^2}~\hat{\mathbf{n}}\,.$ For shells, the initial condition for $\vec{X}(\sigma, t)$ at $t=0$ is chosen as ($r:$ radial coordinate, $k:$~wave number)
\begin{equation}
\label{eq:Sec32}
\vec{\mathbf{X}}_i (\sigma, r) \equiv \vec{\mathbf{X}} (\sigma, t=0)= \left\{-r \cos(2\pi \sigma)\,, r \sin(2\pi \sigma)\,, \delta \sin(2\pi k \sigma)\right\}\,,\quad \sigma\in\left[0\,, 1/k\right]\,.
\end{equation} 
The derivatives of Eq.~\ref{eq:Sec32} with respect to $\sigma$ and $r$ yield tangent vectors to the surface $\partial \vec{\mathbf{X}}/\partial \sigma$ and $\hat{\mathbf{n}}\,.$ By using higher order derivatives and other relations from differential geometry~\cite{Stoker}, the initial conditions for the variables $g\,, \kappa_g\,, \kappa_N\,, \kappa_{N,2}\,, \tau_g\,, U\,,$ the position vector $\vec{\mathbf{X}}\,,$ and the orthonormal triad $\{\partial{\vec{\mathbf{X}}}/\partial s\,, \hat{\mathbf{n}}\,, \hat{\mathbf{N}}\}$ are determined from Eq.~\ref{eq:Sec31} (at $z=0$) and from Eq.~\ref{eq:Sec32} (at $r=1$).

\begin{table*}[!ht]
\begin{center}
\small{\begin{tabular}{|c||c|c|}
 \hline
   Variable & Equation &\\
   \hline\hline
   $\kappa_N$ & $\frac{\partial \kappa_N}{\partial t}=\frac{\partial}{\partial s} (U \tau_g)+\tau_g \frac{\partial U}{\partial s}+\kappa_g (\kappa_N-\kappa_{N, 2})U\,,$ & Eq.~\ref{CMeq2c} \\ \hline
   $\tau_g$ & $\frac{\partial \tau_g}{\partial t}=\frac{\partial}{\partial s}\left(\kappa_{N, 2} U\right)-\kappa_{N}\frac{\partial U}{\partial s}+2 \kappa_g\tau_g U\,,$ & Eq.~\ref{CMeq2d}\\ \hline
   $\sqrt{g}$ & $\frac{\partial \sqrt{g}}{\partial t}=-\sqrt{g}\kappa_g U\,,$ & Eq.~\ref{CMeq9} \\ \hline
   $\kappa_g$ & $\frac{\partial \kappa_g}{\partial t}=\frac{\partial^2 U}{\partial s^2}+\left(\kappa_g^2+\kappa_G\right)U\,,$ & Eq.~\ref{CMeq11}\\ \hline
   $U$ & $U=-\alpha_1 \kappa_g+\alpha_2 \kappa_g^2+\alpha_3 \kappa_g^3+\eta_1 H^2$ & \\
    & $+\eta_{21} \kappa_N \kappa_{N,2}+\eta_{22}\tau_g^2+\eta_3\kappa_g H^2-\eta_{41}\kappa_g \kappa_N \kappa_{N,2}$  &  Eq.~\ref{eq:Sec21}\\
    & $+\eta_{42}\kappa_g\tau_g^2+\eta_5\tau_g H+\eta_6\kappa_g\tau_gH-\lambda \frac{\partial^2 \kappa_g}{\partial s^2}\,.$ & \\ \hline   
    $H\equiv \frac{1}{2}\left(\kappa_{N}+\kappa_{N,2}\right)$ & $\frac{\partial H}{\partial t}=\zeta U H$ & Eq.~\ref{eq:Sec22} \\ \hline
\end{tabular}}
\vskip 0.3cm
\small{\begin{tabular}{|c||c|c|}
 \hline
   Structure & Position & Boundary condition \\
   \hline\hline
   
   All & $\sigma=0\,,$ & $\partial \kappa_g/\partial s=\partial U/\partial s=0\,,$ \\ \hline
   All & $\sigma=1/$k$\,,$ & $\partial \kappa_g/\partial s= \partial U/\partial s=0\,.$ \\ \hline
   \end{tabular}}
   \vskip 0.3cm
\small{\begin{tabular}{|c||c|c|}
 \hline
   Structure & Parameters & Equation\\
   \hline\hline
   Vases, shells & $\alpha_1=1\,,$ $\alpha_2=0.5\,,$ $\alpha_3=1\,,$  $\eta_1=1\,,$ & Eq.~\ref{eq:Sec21} \\ 
     &$\eta_{21}=1\,,$ $\eta_{22}=-1\,,$ $\eta_3=1\,,$ $\eta_{41}=3\,,$  & \\ 
     & $\eta_{42}=3\,, \eta_5=\eta_6=0\,,$ and $\lambda=1\,.$ & \\ \hline
   Vases, shells & $\zeta=-0.4$ & Eq.~\ref{eq:Sec22}\\ \hline
   Vases & $\delta=0.05\,,$ $m^2=0.1\,,$ $k=4$ or $k=6\,.$ & Eq.~\ref{eq:Sec31}\\ \hline
   Shells & $\delta=0.01\,,$  $k=4$ or $k=6\,.$ & Eq.~\ref{eq:Sec32} \\\hline
   \end{tabular}}
   \vskip 0.3cm
\small{\begin{tabular}{|c||c|c|}
   \hline
   Structure & Parameters & Equation\\
   \hline\hline
   Oscillating stems & $\alpha_1=1\,,$ $\alpha_2=1\,,$ $\alpha_3=0.5\,,$  $\eta_1=1.6\,,$ & Eq.~\ref{eq:Sec21} \\ 
    &$\eta_{21}=-0.3\,,$ $\eta_{22}=0.5\,,$  $\eta_3=\eta_{41}=0$ &\\ & $\eta_{42}=\eta_5=\eta_6=0\,,$ and $\lambda=1\,.$ &  \\
    \hline
    Oscillating stems & $\zeta=0.02$ & Eq.~\ref{eq:Sec22}\\ \hline
    Oscillating stems & $\delta=0.1\,,$  $m^2=0.1\,,$ $k=1\,.$ & Eq.~\ref{eq:Sec31}\\ \hline
    \end{tabular}}
\end{center}
\caption{{\bf Equations of motion of the geometrical variables, boundary conditions, and simulation parameters}. The non-linear partial and ordinary differential equations governing the motion of the curvilinear growth site, boundary conditions for each of the simulated shapes shown in Figs. 3-7, and the corresponding simulation parameters are listed. The parameter $k$ is the wave number of the initial perturbation (e.g. $k=4$ for a $4$-fold vase or $k=6$ for a $6$-fold vase.) The Gaussian curvature of the surface at the boundary curve $\kappa_G$ is defined as $\kappa_G\equiv \kappa_N\kappa_{N, 2}-\tau_g^2\,.$ For initial conditions, see Eqs.~\ref{eq:Sec31},~\ref{eq:Sec32}, and the main text. }
\label{table:simulations}
\end{table*}

To reconstruct the surface from a known time series of the geometrical variables $g\,, \kappa_g\,, \kappa_N\,,$ $\kappa_{N,2}\,, \tau_g\,, U\,,$ we make use of Eq.~\ref{eq:Sec11}, which requires knowing the time evolution of $\hat{\mathbf{n}}=\hat{\mathbf{N}}\times \partial\vec{\mathbf{X}}/\partial s$ (see Fig.~\ref{fig:schematics}~a), i.e., $\partial \hat{\mathbf{n}}/\partial t=\partial \left(\hat{\mathbf{N}}\times \partial\vec{\mathbf{X}}/\partial s\right)/\partial t\,.$ The ordinary differential equations that govern the time derivatives of $\partial\vec{\mathbf{X}}/\partial s$ and $\hat{\mathbf{N}}$ are given by Eqs.~\ref{eq:SI13} and~\ref{eq:SI15}, respectively, which are derived in Appendix~\ref{appA}. Thus, the dynamics and the final form of the surface can be mapped from the space of scalar dependent variables $g\,, \kappa_g\,, \kappa_N\,, \kappa_{N,2}\,,\tau_g\,, U\,,$ to the Euclidean space $\mathbb{R}^3$ by (i) $\partial \vec{\mathbf{X}}/\partial t=\hat{\mathbf{n}} U\,,$ (ii) $\hat{\mathbf{n}}=\hat{\mathbf{N}}\times \partial\vec{\mathbf{X}}/\partial s\,,$ (iii) Eq.~\ref{eq:SI13} and (iv) Eq.~\ref{eq:SI15}\,.

\subsection{Growth and form of vases, shells, and oscillating stems}
\label{sec:morphologies}

\begin{figure}[!ht]
\centering
\includegraphics[width=1\columnwidth, clip=true]{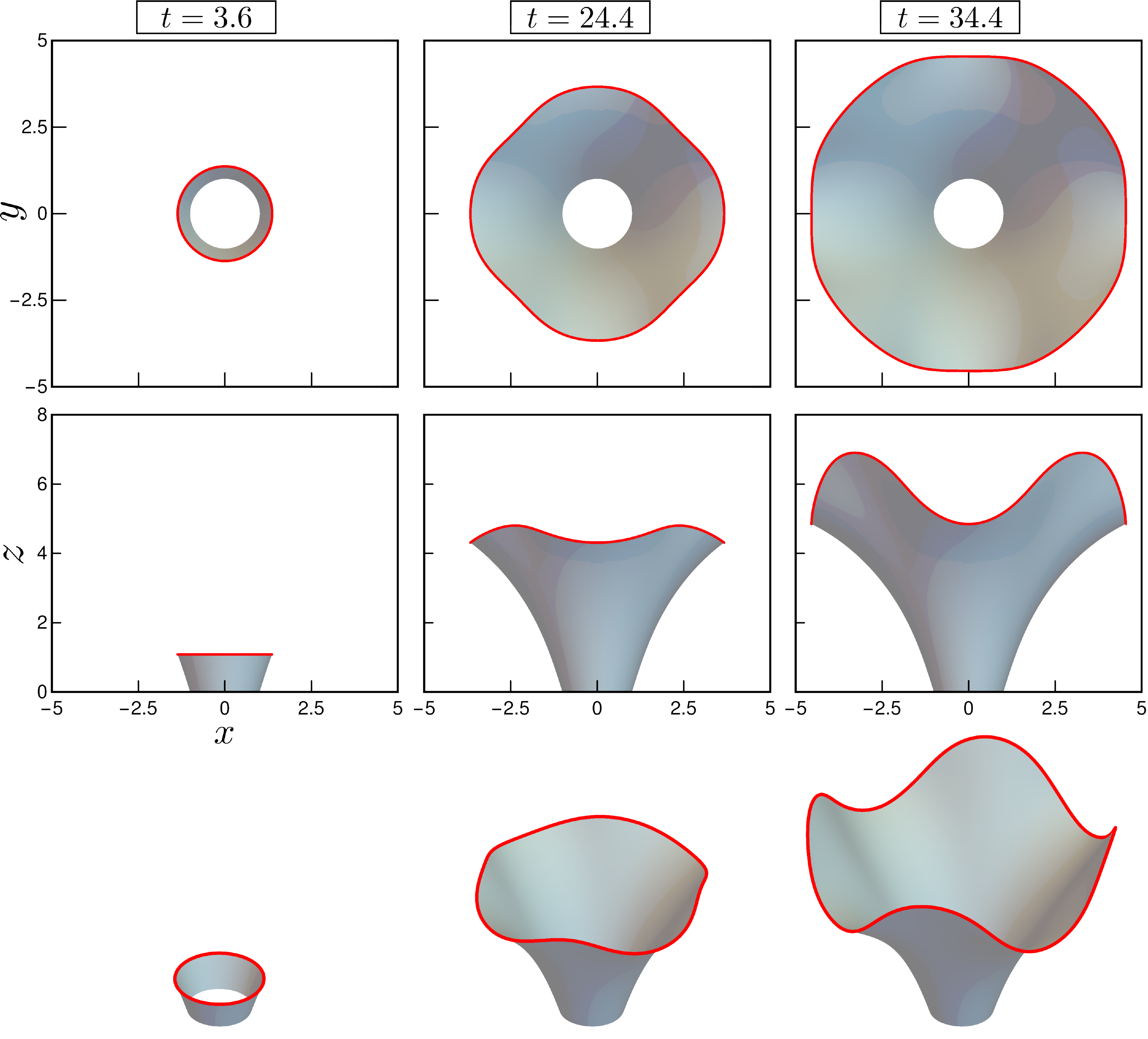}
\caption{{\bf Growth and form of vase-like morphologies with 4-fold symmetry.} For the wave number $k=4$ and the simulation parameters listed in Table~\ref{table:simulations}, the time evolution of the vase growth is presented for three different dimensionless times, $t=3.6\,,$ $t=24.4\,,$ $t=34.4\,.$ The top row shows the plan view, the middle row the side view, and the bottom row the elevated view of the emergence of an undulated vase-like geometry in time. The boundary curve is coloured in red, and the growing surface is coloured in grey. } 
\label{fig:res1}
\end{figure}

\begin{figure}[!ht]
\centering
\includegraphics[width=1\columnwidth, clip=true]{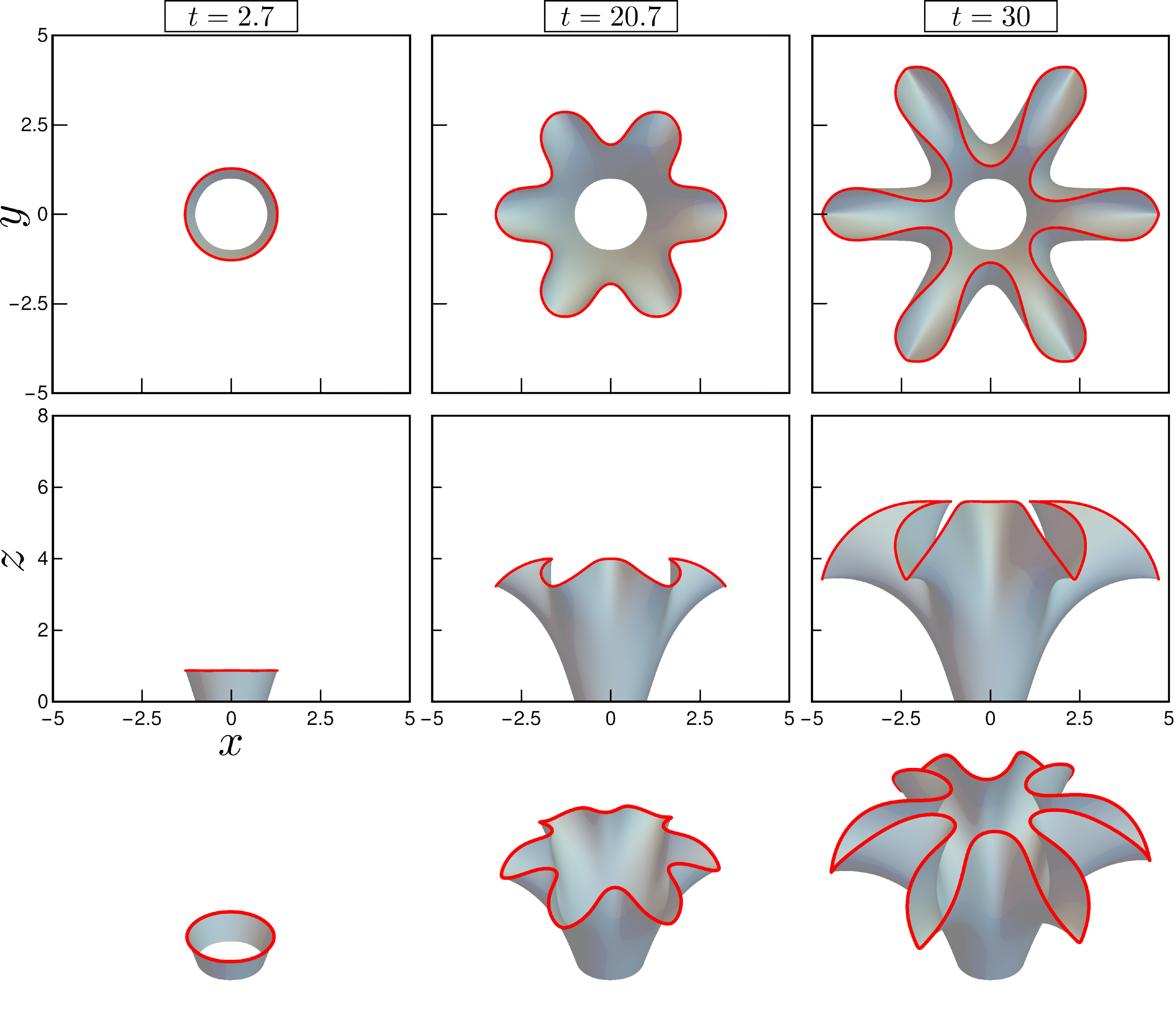}
\caption{{\bf Growth and form of vase-like morphologies with 6-fold symmetry.} For the wave number $k=6$ and the simulation parameters listed in Table~\ref{table:simulations}, the time evolution of the vase growth is presented for three different dimensionless times, $t=2.7\,,$ $t=20.7\,,$ $t=30\,.$ The top row shows the plan view, the middle row the side view, and the bottom row the elevated view of the emergence of an undulated vase-like geometry in time. The boundary curve is coloured in red, and the growing surface is coloured in grey.} 
\label{fig:res2}
\end{figure}

The three classes of shapes presented here, i.e., vases, shells, and oscillating stems, highlight the versatility of our theory (Table~\ref{table:simulations}) in capturing the growth dynamics and form of arbitrarily complex morphologies. The vases are depicted in Fig.~\ref{fig:res1}, Movies S1-S3 for a wave number $k=4$ and in Fig.~\ref{fig:res2}, Movies S4-S6 for $k=6\,,$ and the oscillating stems in Fig.~\ref{fig:res5}, Movies S13-S15 for $k=1\,.$ Eq.~\ref{eq:Sec31} indicates that, on the $xy-$plane, the vases and oscillating stems start growing from a unit circle (radius $r_0=1$), along which the prefactor $\epsilon m^2$ in Eq.~\ref{eq:Sec31} induces undulating perturbations set by $k\,.$ Note that, in Eq.~\ref{eq:Sec21}, the term $\lambda \partial^2 \kappa_g/\partial s^2$ penalizes the local non-uniformity in $\kappa_g$ along the boundary. Yet, while vases grow, because $\lambda=1$ (Table~\ref{table:simulations}) and the total length of the boundary curve $L_c$ becomes much bigger than $r_0=1$ ($L_c\gg 1$), this third-order term becomes insignificant, enabling more pronounced undulations associated with $\kappa_g$ in the local tangent plane to the surface (the $\partial \vec{\mathbf{X}}/\partial s-\hat{\mathbf{n}}$ plane). An oscillating profile of $\kappa_g$ can further induce variations in $\kappa_N$ along the boundary curve through Eq.~\ref{CMeq2c}, as shown in Fig.~\ref{fig:res2}, Movies S4-S6. All these effects are largely suppressed for a boundary curve that remains sufficiently short ($L_c\sim 1$) as observed in the growth of the oscillating stems (Fig.~\ref{fig:res5}, Movies S13-S15). Yet, because $\zeta<0$ in Eq.~4 (Table~\ref{table:simulations}), the mean curvature $H$ decreases over time from its positive value at $t=0\,,$ Then, since $\kappa_N\geq 0$ at the interface at all times, $\kappa_{N,2}$ changes sign to satisfy a low mean curvature and subsequently becomes positive again. This temporal alternation yields an oscillating structure for the set of parameters of the growth speed $U$ listed in Table~\ref{table:simulations}; $U$ eventually vanishes after $5$ periods, terminating growth (Fig.~\ref{fig:res5}, Movies S13-S15).

\begin{figure}[!ht]
\centering
\includegraphics[width=1\columnwidth, clip=true]{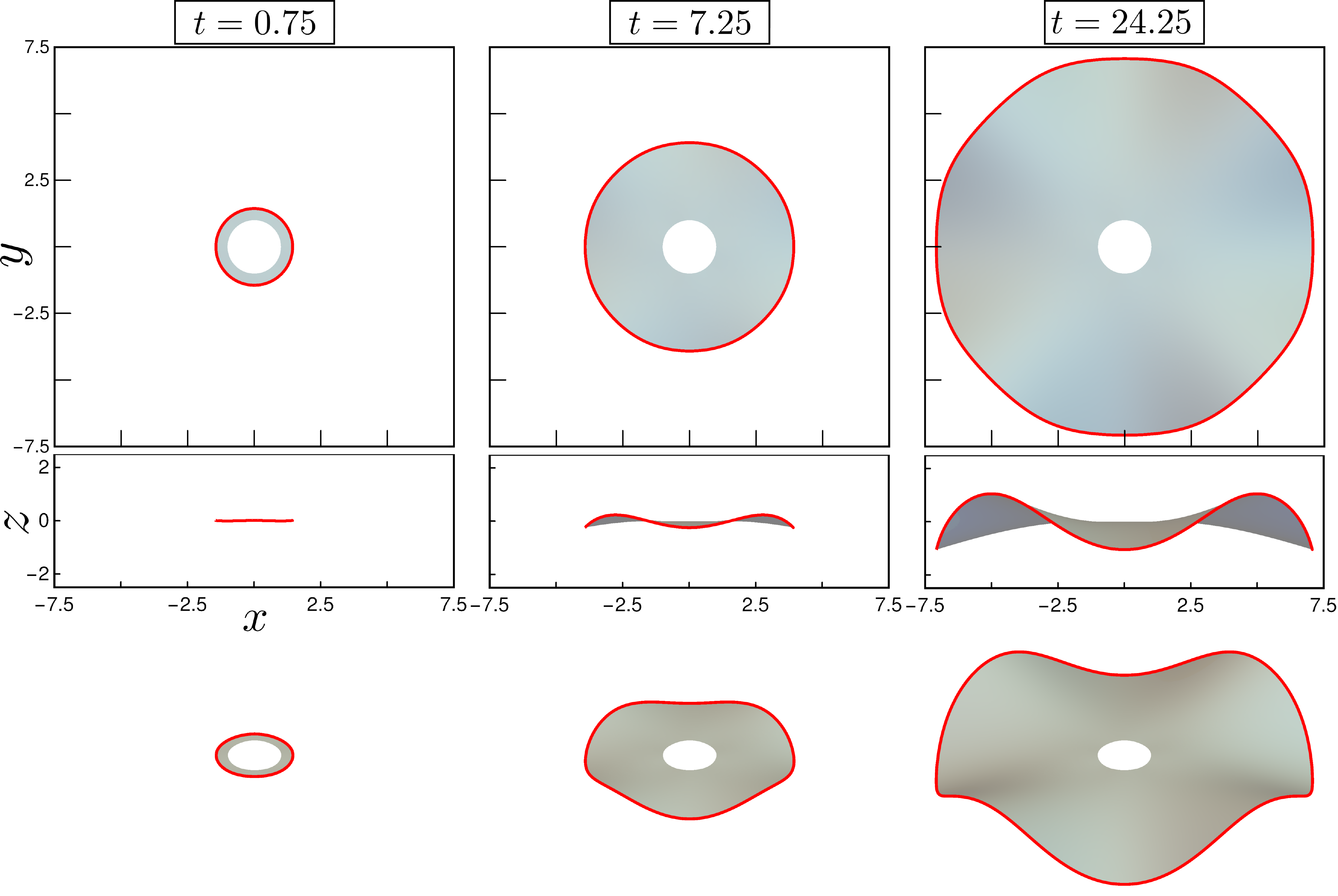}
\caption{{\bf Growth and form of shell-like morphologies with 4-fold symmetry.} For the wave number $k=4$ and the simulation parameters listed in Table~\ref{table:simulations}, the time evolution of the shell growth is presented for three different dimensionless times, $t=0.75\,,$ $t=7.25\,,$ $t=24.25\,.$ The top row shows the plan view, the middle row the side view, and the bottom row the elevated view of the emergence of an undulated shell-like geometry in time. The boundary curve is coloured in red, and the growing surface is coloured in grey.} 
\label{fig:res3}
\end{figure}

\begin{figure}[!ht]
\centering
\includegraphics[width=1\columnwidth, clip=true]{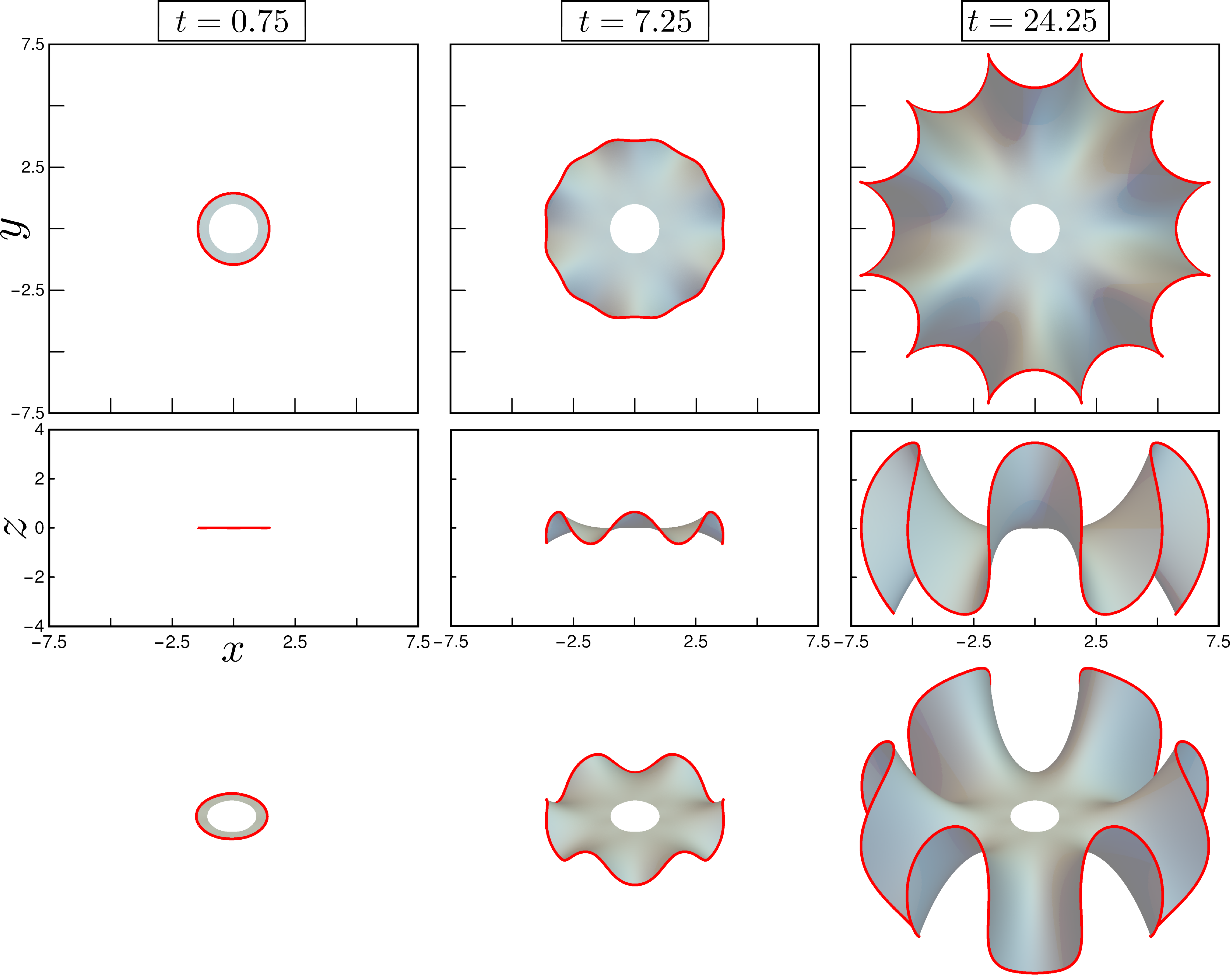}
\caption{{\bf Growth and form of shell-like morphologies with 6-fold symmetry.} For the wave number $k=6$ and the simulation parameters listed in Table~\ref{table:simulations}, the time evolution of the shell growth is presented for three different dimensionless times, $t=0.75\,,$ $t=7.25\,,$ $t=24.25\,.$ The top row shows the plan view, the middle row the side view, and the bottom row the elevated view of the emergence of an undulated shell-like geometry in time. The boundary curve is coloured in red, and the growing surface is coloured in grey.} 
\label{fig:res4}
\end{figure}

\begin{figure}[!ht]
\centering
\includegraphics[height=0.8\textheight, clip=true]{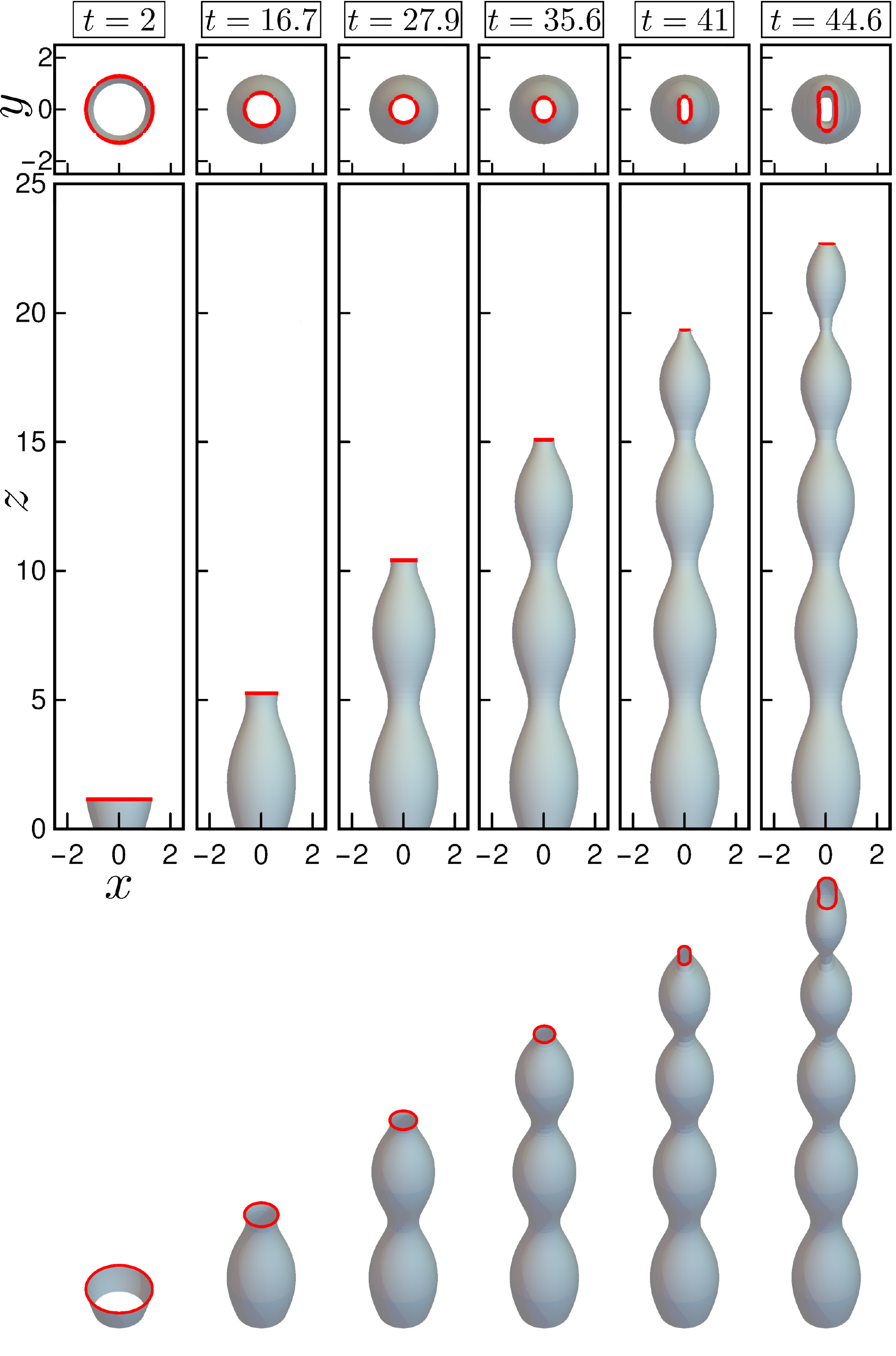}
\caption{{\bf Growth and form of oscillating stem-like morphologies.} For the simulation parameters listed in Table~\ref{table:simulations}, the time evolution of the oscillating stem growth is presented for five different dimensionless times, $t=2\,,$ $t=16.7\,,$ $t=27.9\,,$ $t=35.6\,,$ $t=41\,,$ $t=44.6\,.$ The top row shows the plan view, the middle row the side view, and the bottom row the elevated view of the emergence of an oscillating stem geometry in time. The boundary curve is coloured in red, and the growing surface is coloured in grey.}
\label{fig:res5}
\end{figure}

The growth and form of shells is presented in Fig.~\ref{fig:res3}, Movies S7-S9 for $k=4$ and in Fig.~\ref{fig:res4}, Movies S10-S12 for $k=6\,.$ Based on Eq.~\ref{eq:Sec32} the shells start growing from a unit semicircle, and the prefactor $\delta$ in Eq.~\ref{eq:Sec32} induces undulations set by the wave number $k\,.$ As with the vases, these undulations are amplified in the $\partial \vec{\mathbf{X}}/\partial s-\hat{\mathbf{n}}$ plane for $\lambda=1$ while the shells grow, inducing oscillations in $\kappa_g$ through Eqs.~\ref{CMeq11} and~\ref{eq:Sec21}, and in turn in $\kappa_N$ through Eq.~\ref{CMeq2c}. The emergence of ripples in the $\hat{\mathbf{N}}-\partial \vec{\mathbf{X}}/\partial s$ plane is mainly due to the fact that the initial mean curvature satisfies $|H\big|_{t=0}|\ll1\,,$ and for $\zeta=0.02$ (see Table~\ref{table:simulations}) then $|H|$ remains sufficiently low throughout the entire simulation time $t_{total}=25\,.$ Then, as the normal curvature $\kappa_N$ increases, $\kappa_{N, 2}$ will also increase, albeit with an opposite sign, to satisfy a low mean curvature.

\section{Discussion and conclusions}
\label{sec:discussion}

Accretionary growth through mineralization of natural solid composites, such as molluscan and brachiopod shells~\cite{shells1, shells2, shells3, shells4, shells5}, and of analogous chemical precipitates, such as complex carbonate-silica patterns~\cite{Ruiz1, Ruiz3, Wim, Kellermeier} or chemical garden tubes~\cite{Gardens1, Gardens2}, involves an emerging high-aspect-ratio wall that can be approximated as a 2D smooth surface where growth occurs through the dynamics of a space curve that coarse-grains the narrow reaction front. To that end, we have presented a general geometrical theory of growth and form of a 2D surface at its margin that takes into account the scale-free geometric constraints based on the theory of surfaces~\cite{Stoker}. Our framework captures the formation of any non-planar smooth surface when complementing the scale-free geometric constraints, i.e., the Codazzi-Mainardi equations (Eqs.~\ref{CMeq2c} and~\ref{CMeq2d}), the Gauss {\it theorema egregium} (Eq.~\ref{CMeq11}), and the evolution of the curve metric (Eq.~\ref{CMeq9}) by two physical-chemical closure relations for the growth speed and local curling dynamics specific to a given system. With the goal of simulating hypothetical and aesthetic morphologies, we have proposed two simple closure relations: (i) The expansion of the growth speed $U$ into a power series of scalar geometric variables where each term obeys two reversal symmetries in the surface normal and the local arc length coordinate, $\hat{\mathbf{N}}\rightarrow -\hat{\mathbf{N}}$ and $s\rightarrow -s$ (Eq.~\ref{eq:Sec21}), (ii) a first-order ordinary differential equation for the time evolution of the mean curvature $H$ and thereby the second normal curvature $\kappa_{N, 2}$ (Eq.~\ref{eq:Sec22}). The simulated morphologies, although being abstract, closely resemble natural patterns: the form and perodicity of shell-like shapes are similar to those of molluscan and brachiopod shells, and oscillating-stem like shapes look like the solid tubes observed in chemical gardens.

Another relevance of our theory for the propagation of curves on non-planar surfaces is to a potential generalization of the Korteweg-de Vries (KdV) and modified Korteweg-de Vries (mKdV) hierarchies of integrable systems~\cite{KdV, Darboux}. These hierarchies return the celebrated KdV and mKdV equations that explain the existence of solitary waves and their propagation dynamics~\cite{Darboux}. The mKdV hierarchy can be shown to be equivalent to the dynamics of a closed contour on a plane when conservation of the perimeter of the curve and the area enclosed by it are imposed. The main assumption to construct the mKdV hierarchy is that the equation of motion of a closed curve constrained to a plane is given by $\partial \vec{\mathbf{X}}/\partial t=\hat{\mathbf{n}}U+ (\partial \vec{\mathbf{X}}/\partial s) V\,,$ where the tangent speed $V$ is taken to be periodic over the curve perimeter.  Noting that the total curvature $\kappa$ of a planar curve satisfies $\kappa=\kappa_g\,,$ the perimeter and area are conserved when $U$ and $\kappa U$ are total derivatives of a periodic function with respect to the arc length variable $s$~\cite{KdV}. Construction of the mKdV hierarchy further yields an infinite number of conserved quantities, which are given by the even-order polynomials of the curvature and its derivatives with respect to $s\,.$ The mKdV hierarchy is linked to the KdV hierarchy for a complex variable $\omega(s, t)$ through the Miura transformation, which defines the real and imaginary parts of $\omega(s, t)$ in terms of $\kappa$ and its derivatives~\cite{Miura}. One necessary condition to build these hierarchies is that the growth rate $U$ must be chiral, i.e., it must break the $s\rightarrow -s$ symmetry along the curve. In our model, relaxing this symmetry when seeking a form for $U(\kappa_g, \kappa_N, \tau_g, \kappa_{N, 2})$ and an accompanying closure for $\kappa_{N,2}$ may allow for the construction of broader mathematical hierarchies yielding a set of non-linear partial differential equations (thus in general being non-integrable) rather than one equation (e.g. mKdV equation in the mKdV hierarchy). This mathematical endeavor may have broader physical implications ranging from the soliton dynamics on curved surfaces to incompressible hydrodynamic flows.

Despite the strengths presented here, currently, our theory has several limitations. The primary problem is the rigorous determination of physical closure relations: For natural shell growth, biomineralization couples transport and reaction of species and their solidification at the growth front, where the length scales of the steep concentration gradients and the structure are separated. Inorganic model systems, such as carbonate-silica precipitates and chemical gardens, exhibit analogous dynamics. The challenge lies at resolving the scale separation and non-locality of species transport, as well as determining the physical laws that enable steering the gradients to guide the position, direction, and local ordering of assembly. To that end, detailed experiments that characterize the system at the time scale of growth and the thickness scale of the interface are needed. The subsequent microscopic or continuum-level theories, which would need to be developed to explain the pertinent experimental data, can then be coarse-grained to physical closure laws, which can replace the closure relations presented here, liberating our theory from an extensive set of free parameters.

A second challenge is imposing steric repulsions between initially distant wall sections to avoid intersection at a later time during growth. Indeed, the structures presented in Figs.~\ref{fig:res1},~\ref{fig:res2},~\ref{fig:res4} are naturally not self-avoiding for longer simulation times or for sets of parameters different than in Table~\ref{table:simulations}. To overcome that, the necessary long-range interactions can only be introduced through the closure relations. If the growth is diffusion limited as e.g. measured in carbonate-silica co-precipitation~\cite{Kaplan}, diffusion of chemical species around the interface in the background fluid would yield a growth rate proportional to the diffusive flux~\cite{Langer, MullinsSekerka}, which diminishes when two wall sections come very close to each other, thereby locally suppressing growth. Another way to implement self-avoidance is considering the "nematic" long-range order of parallel vector fields (along two families of curves, each locally parallel to $\partial\vec{\mathbf{X}}/\partial s$ or $\hat{\mathbf{n}}$) on the growing non-planar surface and penalizing the "isotropic" phase corresponding to the defects in nematic ordering, which would occur at the intersection sites. Technically, this would require closing the purely geometric equations (Eqs.~\ref{CMeq9},~\ref{CMeq2c},~\ref{CMeq2d}, and~\ref{CMeq11}) with the Euler-Lagrange (EL) equations (and their overdamped dynamics) for the minimization of the Landau-de Gennes (LdG) free energy of liquid crystals~\cite{deGennes, BluePhases}, in which the nematic director field must involve parallel transport of vectors on the surface in the sense of Levi-Civita~\cite{Stoker}. The technical difficulty here is that Eqs.~\ref{CMeq9},~\ref{CMeq2c},~\ref{CMeq2d}, and~\ref{CMeq11} are in a Lagrangian frame, whereas the EL equations of the LdG free energy must be evaluated in an Eulerian frame. An Eulerian description of planar curve motion was previously developed~\cite{KdV} and must ideally be generalized to motion on non-planar smooth surfaces.

These limitations notwithstanding, by writing the surface differential-geometric compatibility equations in a dynamical setting, along with two closure relations for the growth speed and curling rate, we have developed a geometric theory of edge-driven growth of a smooth, simply connected non-Euclidean surface embedded in three dimensions. Simulations of the governing equations with appropriate initial and boundary conditions lead to morphologies that resemble a variety of natural and artificial precipitating thin-walled structures. When complemented by experimentally determined parameters in the symmetry-based closure relations, our theory has the potential to provide a quantitative theoretical understanding that paves the way for harnessing self-assembly processes to engineer complex morphologies with tailored material properties.

\paragraph*{Data Accessibility.} The Supplementary Movies S1-S15 for the simulations of flower-like, shell-like, and oscillating stem-like morphologies can be downloaded at XXXX. The Python source codes of the simulations can be downloaded at \url{https://github.com/nadirkaplan/geometrically_constrained_growth}.

\paragraph*{Authors' Contributions.} C.N.K. and L.M. conceived the mathematical model, interpreted the theoretical and computational results and wrote the paper. C.N.K. derived the model, implemented and performed the simulations in consultation with L.M.

\paragraph*{Competing Interests.} The authors declare no conflicts of interest.

\paragraph*{Funding.} We thank the College of Science at Virginia Tech (CNK), and the US NSF grants DMR-2011754 and DMR-1922321 (LM) as well as the Henri Seydoux Fund for partial financial support.

\paragraph*{Acknowledgements.} The authors thank W.~L. Noorduin and J.~Aizenberg for fruitful discussions.

\vskip6pt

\renewcommand{\theequation}{A\arabic{equation}}
\setcounter{equation}{0}

\section*{Appendix A: Time evolution of the orthonormal triad}
\label{appA}

Here, we derive the time derivatives of the vectors $\hat{\mathbf{n}}\,,$ $\hat{\mathbf{N}}\,,$ and $\partial \vec{\mathbf{X}}/\partial s\,.$ In the simulations, these auxiliary equations are needed to map the time evolution the surface growth from the space of scalar dependent variables $g\,, \kappa_g\,, \kappa_N\,, \tau_g\,, \kappa_{N,2}\,, U\,,$ to the Euclidean space $\mathbb{R}^3\,.$

As a first preliminary relation, the combination of Eqs.~\ref{eq:CMeq00},~\ref{CMeq9} and the relation $\sqrt{g}\partial/\partial s=\partial/\partial\sigma$ result in the following identity between the mixed derivatives,
\begin{equation}
\label{eq:SI5}
\frac{\partial}{\partial t}\frac{\partial}{\partial s}=\frac{\partial}{\partial s}\frac{\partial}{\partial t}+\kappa_g U \frac{\partial}{\partial s}\,.
\end{equation}
A second preliminary relation expresses $\partial \mathbf{\hat{n}}/\partial s$ in terms of its components along $\mathbf{\hat{n}}$ and $\partial \vec{\mathbf{X}}/\partial s$ since $\partial \mathbf{\hat{n}}/\partial s\perp \mathbf{\hat{n}}$ based on Eq.~\ref{eq:SI7}. We calculate $\partial \mathbf{\hat{n}}/\partial s$ by taking the derivative of $\mathbf{\hat{n}}= \mathbf{\hat{N}}\times\partial \vec{\mathbf{X}}/\partial s\,,$ which becomes
\begin{equation}
\label{eq:SI9}
\frac{\partial \mathbf{\hat{n}}}{\partial s}=\frac{\partial \mathbf{\hat{N}}}{\partial s}\times\frac{\partial \vec{\mathbf{X}}}{\partial s}-\kappa_g\frac{\partial \vec{\mathbf{X}}}{\partial s}\,,
\end{equation}
The second term on the right-hand side of Eq.~\ref{eq:SI9} is obtained by evaluating $\mathbf{\hat{N}}\times \partial^2 \vec{\mathbf{X}}/\partial s^2\,,$ where
\begin{equation}
\label{eq:SI8}
\frac{\partial^2 \vec{\mathbf{X}}}{\partial s^2}=\kappa_N \mathbf{\hat{N}}+\kappa_g\mathbf{\hat{n}}
\end{equation}
by using the definitions of $\kappa_g$ and $\kappa_N$ in Table~\ref{table:geometry}. Since $\partial\mathbf{\hat{N}}/\partial s$ should lie in the plane spanned by $\mathbf{\hat{n}}$ and $\partial \mathbf{\vec{X}}/\partial s\,,$ the first term  on the right-hand side of Eq.~\ref{eq:SI9} must be either parallel or anti-parallel to $\mathbf{\hat{N}}\,.$ Then, by using the definition of the geodesic torsion (Table~\ref{table:geometry}), Eq.~\ref{eq:SI9} becomes
\begin{equation}
\label{eq:SI10}
\frac{\partial \mathbf{\hat{n}}}{\partial s}=\tau_g \mathbf{\hat{N}}-\kappa_g \frac{\partial \vec{\mathbf{X}}}{\partial s}\,.
\end{equation}

By virtue of the two preliminary relations Eqs.~\ref{eq:SI5} and~\ref{eq:SI10}, we can now calculate the time derivatives of the unit vectors. We first evaluate $\partial (\partial \vec{\mathbf{X}}/\partial s)/\partial t$ by implementing Eq.~\ref{eq:SI5}:
\begin{equation}
\label{eq:SI12}
\frac{\partial}{\partial t}\frac{\partial \vec{\mathbf{X}}}{\partial s}=\frac{\partial}{\partial s}\frac{\partial \vec{\mathbf{X}}}{\partial t}+\kappa_g U \frac{\partial \vec{\mathbf{X}}}{\partial s}\,.
\end{equation}
When $\partial(\partial \vec{\mathbf{X}}/\partial t)/\partial s$ is evaluated by using Eq.~\ref{eq:Sec11} and Eq.~\ref{eq:SI10}, then Eq.~\ref{eq:SI12} becomes
\begin{equation}
\label{eq:SI13}
\frac{\partial}{\partial t}\frac{\partial \vec{\mathbf{X}}}{\partial s}=U\tau_g\mathbf{\hat{N}}+\frac{\partial U}{\partial s} \mathbf{\hat{n}}\,.
\end{equation}

The definition of the second normal curvature $\kappa_{N, 2}$ (Table~\ref{table:geometry}) allows us to determine one of the components of $\partial\mathbf{\hat{n}}/\partial t\,.$ Its second component can be extracted by dotting $\mathbf{\hat{n}}$ into Eq.~\ref{eq:SI13} and using the product rule subsequently. These steps yield
\begin{equation}
\label{eq:SI14}
\frac{\partial\mathbf{\hat{n}}}{\partial t}=U\kappa_{N, 2} \mathbf{\hat{N}}-\frac{\partial U}{\partial s} \frac{\partial \vec{\mathbf{X}}}{\partial s}\,.
\end{equation}
Taking the scalar product of $\mathbf{\hat{N}}$ with Eqs.~\ref{eq:SI13} and~\ref{eq:SI14} and applying the product rule gives $\partial \mathbf{\hat{N}}/\partial t$ as
\begin{equation}
\label{eq:SI15}
\frac{\partial \mathbf{\hat{N}}}{\partial t}=-U\kappa_{N, 2} \mathbf{\hat{n}}-U \tau_g \frac{\partial \vec{\mathbf{X}}}{\partial s}\,.
\end{equation}

\end{document}